\documentclass[journal=nalefd,manuscript=letter]{achemso}
\usepackage[version=3]{mhchem} 
\usepackage{mathpazo}
\usepackage{amsmath}
\usepackage{lmodern}
\usepackage{caption,geometry}
\usepackage{color}
\usepackage{soul}

\usepackage{amssymb,amsmath,graphicx,lineno,hyperref,mciteplus}

\title{Dynamic acoustic control of individual optically active quantum dot-like emission centers in heterostructure nanowires}

\author{Matthias Wei\ss}
\affiliation[Univ. Augsburg]{Lehrstuhl f\"{u}r Experimentalphysik 1 and Augsburg Centre for Innovative Technologies ({\it ACIT}), Universit\"{a}t Augsburg, Universit\"{a}tsstr. 1, 86159 Augsburg, Germany} 

\author{J\"{o}rg B. Kinzel}
\affiliation[Univ. Augsburg]{Lehrstuhl f\"{u}r Experimentalphysik 1 and Augsburg Centre for Innovative Technologies ({\it ACIT}), Universit\"{a}t Augsburg, Universit\"{a}tsstr. 1, 86159 Augsburg, Germany} 
\alsoaffiliation[NIM]{Nanosystems Initiative Munich, Schellingstr. 4, 80799 M\"{u}nchen, Germany}

\author{Florian J. R. Sch\"ulein}
\affiliation[Univ. Augsburg]{Lehrstuhl f\"{u}r Experimentalphysik 1 and Augsburg Centre for Innovative Technologies ({\it ACIT}), Universit\"{a}t Augsburg, Universit\"{a}tsstr. 1, 86159 Augsburg, Germany} 
\alsoaffiliation[NIM]{Nanosystems Initiative Munich, Schellingstr. 4, 80799 M\"{u}nchen, Germany}

\author{Michael Heigl}
\affiliation[Univ. Augsburg]{Lehrstuhl f\"{u}r Experimentalphysik 1 and Augsburg Centre for Innovative Technologies ({\it ACIT}), Universit\"{a}t Augsburg, Universit\"{a}tsstr. 1, 86159 Augsburg, Germany} 

\author{Daniel Rudolph}
\affiliation[WSI, TU M\"unchen]{Walter Schottky Institut and Physik Department, Technische Universit\"at M\"unchen, Am Coulombwall 4, 85748 Garching, Germany}
\alsoaffiliation[NIM]{Nanosystems Initiative Munich, Schellingstr. 4, 80799 M\"{u}nchen, Germany}

\author{Stefanie Mork\"{o}tter}
\affiliation[WSI, TU M\"unchen]{Walter Schottky Institut and Physik Department, Technische Universit\"at M\"unchen, Am Coulombwall 4, 85748 Garching, Germany}

\author{Markus D\"{o}blinger}
\affiliation[Chemistry, LMU]{Department of Chemistry, Ludwig-Maximilians-Universit\"{a}t M\"{u}nchen, 81377 M\"{u}nchen, Germany}
\alsoaffiliation[NIM]{Nanosystems Initiative Munich, Schellingstr. 4, 80799 M\"{u}nchen, Germany}
\alsoaffiliation[CeNS]{Center for NanoScience ({\it CeNS}), Ludwig-Maximilians-Universit\"{a}t M\"{u}nchen, Geschwister-Scholl-Platz 1, 80539 M\"{u}nchen, Germany}

\author{Max Bichler}
\affiliation[WSI, TU M\"unchen]{Walter Schottky Institut and Physik Department, Technische Universit\"at M\"unchen, Am Coulombwall 4, 85748 Garching, Germany}

\author{Gerhard Abstreiter}
\affiliation[WSI, TU M\"unchen]{Walter Schottky Institut and Physik Department, Technische Universit\"at M\"unchen, Am Coulombwall 4, 85748 Garching, Germany}
\alsoaffiliation[IAS, TU M\"unchen]{Institute for Advanced Study (IAS), Technische Universit\"at M\"unchen, Lichtenbergstra\ss e 2a, 85748 Garching, Germany}
\alsoaffiliation[NIM]{Nanosystems Initiative Munich, Schellingstr. 4, 80799 M\"{u}nchen, Germany}

\author{Jonathan J. Finley}
\affiliation[WSI, TUM]{Walter Schottky Institut and Physik Department, Technische Universit\"at M\"unchen, Am Coulombwall 4, 85748 Garching, Germany}
\alsoaffiliation[NIM]{Nanosystems Initiative Munich, Schellingstr. 4, 80799 M\"{u}nchen, Germany}

\author{Gregor Koblm\"uller}
\affiliation[WSI, TU M\"unchen]{Walter Schottky Institut and Physik Department, Technische Universit\"at M\"unchen, Am Coulombwall 4, 85748 Garching, Germany}
\alsoaffiliation[NIM]{Nanosystems Initiative Munich, Schellingstr. 4, 80799 M\"{u}nchen, Germany}

\author{Achim Wixforth}
\affiliation[Univ. Augsburg]{Lehrstuhl f\"{u}r Experimentalphysik 1 and Augsburg Centre for Innovative Technologies ({\it ACIT}), Universit\"{a}t Augsburg, Universit\"{a}tsstr. 1, 86159 Augsburg, Germany} 
\alsoaffiliation[NIM]{Nanosystems Initiative Munich, Schellingstr. 4, 80799 M\"{u}nchen, Germany}
\alsoaffiliation[CeNS]{Center for NanoScience ({\it CeNS}), Ludwig-Maximilians-Universit\"{a}t M\"{u}nchen, Geschwister-Scholl-Platz 1, 80539 M\"{u}nchen, Germany}

\author{Hubert J. Krenner}\email{hubert.krenner@physik.uni-augsburg.de}
\affiliation{Lehrstuhl f\"{u}r Experimentalphysik 1 and Augsburg Centre for Innovative Technologies ({\it ACIT}), Universit\"{a}t Augsburg, Universit\"{a}tsstr. 1, 86159 Augsburg, Germany} 
\alsoaffiliation[NIM]{Nanosystems Initiative Munich, Schellingstr. 4, 80799 M\"{u}nchen, Germany}
\alsoaffiliation[CeNS]{Center for NanoScience ({\it CeNS}), Ludwig-Maximilians-Universit\"{a}t M\"{u}nchen, Geschwister-Scholl-Platz 1, 80539 M\"{u}nchen, Germany}

\begin{document}
\newpage
\begin{abstract}
We probe and control the optical properties of emission centers forming in radial heterostructure \ce{GaAs}-\ce{Al_{0.3}Ga_{0.7}As} nanowires and show that these emitters, located in \ce{Al_{0.3}Ga_{0.7}As} layers, can exhibit quantum-dot like characteristics. We employ a radio frequency surface acoustic wave to dynamically control their emission energy and occupancy state on a nanosecond timescale. In the spectral oscillations we identify unambiguous signatures arising from both the mechanical and electrical component of the surface acoustic wave. In addition, different emission lines of a single emission center exhibit pronounced anti-correlated intensity oscillations during the acoustic cycle. These arise from a dynamically triggered carrier extraction out of the emission center to a continuum in the radial heterostructure. Using finite element modeling and Wentzel-Kramers-Brillouin theory we identify quantum tunneling as the underlying mechanism. These simulation results quantitatively reproduce the observed switching and show that in our systems these emission centers are spatially separated from the continuum by $>10.5\,{\rm nm}$.
\end{abstract}



Over the past decades the paradigm of bandstructure engineering \cite{Capasso:87} led to novel quantum- and optoelectronic devices using planar semiconductor heterostructures, quantum wells (QWs)\cite{Dupuis1978,*Faist:94}, quantum wires\cite{Kapon1989} and quantum dots (QDs)\cite{Nomura2010}. More recently, first promising steps towards the implementation of heterostructures on a nanowire (NW) platform have been made and first quantum- and optoelectronic devices \cite{Lauhon2002,*Mata2013,*Hyun2013} have been demonstrated. In this active field of nanotechnology, zero-dimensional QD nanostructures are of particular interest since they provide bright single photon emitters \cite{Borgstrom:05a,*Reimer2012} and significant progress has been made over the past years to tailor their fabrication \cite{Dalacu2012,*Makhonin2013} and to control their quantum confined few particle spectrum\cite{Kouwen:10a,*Reimer2011}. As in conventional, planar heterostructures, a second key capability lies in the precise control of the interactions between multiple QDs\cite{Schedelbeck:97,*Krenner:05b,*Stinaff:06} or between QDs and systems of higher dimensionality\cite{Mazur2010}. In NWs, in addition to the aforementioned axial QDs, radial heterostructure QWs\cite{Morral:08b}, QDs\cite{Uccelli:10a} and combinations of QWs and QDs\cite{Heiss2013a} have been fabricated and characterized in optical experiments.\\

While in most experiments performed on optically active QDs \emph{static} control parameters have been applied, recently first steps have been made to employ radio frequency surface acoustic waves (SAWs) to dynamically control charge carrier dynamics and the occupancy state of QDs on NW and nanotube platforms \cite{Kinzel:11,Hernandez:12,Regler2013}. These works have built on schemes which have been established over the past 15 years for planar heterostructures\cite{Rocke:97,*Wiele:98,*Boedefeld:06,*Couto:09,*Voelk:10b}. The underlying mechanism in these experiments is the spatial dissociation and transport of photogenerated electron-hole (e-h) pairs, excitons by the large electric fields and potential induced by the periodic mechanical deformation in a piezoelectric material. The propagation of the SAW itself regulates the injection of e's and h's giving rise to a precisely timed emission of (quantum) light with low temporal jitter at radio frequencies up to the gigahertz range. So far all experiments on planar and NW-based heterostructures have been limited to acousto-electrically induced transport and carrier injection. However, advanced concepts aim to implement optically active and electrostatically defined QDs on a single NW which crucially require the controlled extraction of single charges from a heterostructure QD.\\

In this letter we report on optical experiments performed on QD-like emission centers (ECs) forming in \ce{Al_{0.3}Ga_{0.7}As} layers of radial heterostructure \ce{GaAs}-\ce{Al_{0.3}Ga_{0.7}As} NWs which are coupled to the 2D and 3D continuum of states of a radial QW and the NW core, respectively. We show that the emission can exhibit QD-like properties and apply a SAW control to these nanostructures. In our SAW experiments we resolve clear spectral and anti-correlated intensity oscillations between different EC emission lines. The spectral oscillations are a superposition of dynamic strain-driven deformation potential couplings and electric field-driven Stark-effect tuning. Due to the unique energetics of our structure, we can unambiguously attribute the anti-correlated intensity oscillations to dynamically-modulated carrier tunneling out of the EC into a continuum of higher dimensionality. This first time observation of such mechanism is found to be in quantitative agreement with the calculated efficiency of this process. Furthermore, our modeling predicts for our structure that these ECs have to be spatially separated from a continuum of states by at least $10.5\,{\rm nm}$.\\

\begin{figure}[htb]
	\begin{center}
		\includegraphics[width=0.85\columnwidth]{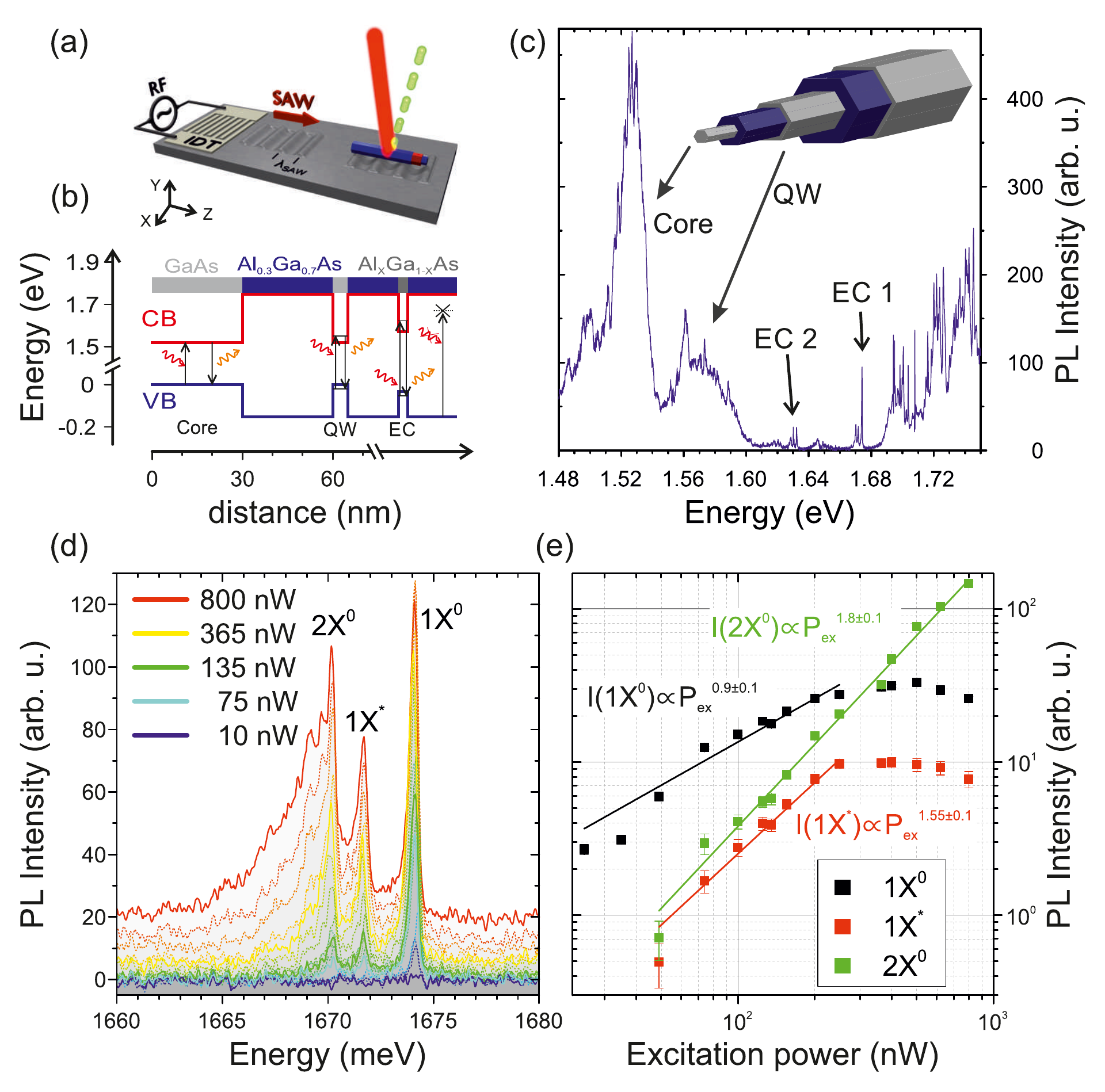}
		\caption{\textbf{Sample, bandstructure and optical characterization} -- (a) Schematic of hybrid NW-SAW chip device. (b) Bandstructure of radial heterostructure and optical pumping (up arrows) and emission processes (down arrows) marked for core, QW and EC. The laser energy does not allow for photogeneration in the \ce{Al_{0.3}Ga_{0.7}As} barriers. (c) Overview PL spectrum of a single NW. The origin of the different signal contributions are labeled and indicated by the schematic of the radial heterostructure NW. (d) Optical pump power dependent PL spectra of EC1 showing a characteristic multi-exciton generation. (e) Extracted peak intensities of the three dominant emission lines as a function of optical pump power in double-logarithmic representation reveals characteristic power-law dependences for neutral single $(1X^0)$, biexciton $(2X^0)$ and a charged exciton $(1X^*)$.}
		\label{Fig:1}
	\end{center}
\end{figure}

The investigated NWs were grown by molecular beam epitaxy (MBE) in a Ga-assisted autocatalytic growth process on a silicon substrate\cite{Rudolph2011}. Under the selected growth conditions these NWs are predominantly of zinc blende (ZB) crystal structure with occasional twin defects and have lengths $l_{NW}>10\,\mu \mathrm{m}$. In the radial direction the as-grown NWs consist of a 60\,nm diameter GaAs core capped by a 100 nm thick \ce{Al_{0.3}Ga_{0.7}As} shell. Within this shell we included a 5\,nm thick radial GaAs quantum well (QW) at a distance of 30\,nm from the core. For passivation the wires are coated by a 5\,nm thick capping layer of GaAs to protect the NWs against oxidation. Details regarding the growth of this complex core-shell NW structure can be found elsewhere\cite{Rudolph2013}. The energy band profile of this radial heterostructure is shown in Figure \ref{Fig:1}(b). For our acoustic measurements we mechanically transferred the NWs onto a YZ-cut LiNbO$_3$ substrate with lithographically defined interdigital transducers (IDTs) for SAW excitation. By applying a RF signal to the IDT a Rayleigh-type SAW is excited which propagates on a Y-cut \ce{LiNbO3} substrate along the Z-direction. The design of the IDTs in this case allows for the excitation of SAWs with a wavelength of $\lambda_{\rm SAW}=18\mu\mathrm{m}$, corresponding to a resonance frequency of $f_{\rm SAW}=\omega_{\rm SAW}/2\pi=194 \, \mathrm{MHz}$ and acoustic period $T_{\rm SAW} = 5.15 \,{\rm ns}$. NWs are transferred from suspension directly onto the SAW-chip\cite{Kinzel:11}. After transfer, we selected NWs with their $(111)$ growth axis oriented within $\pm5^{\rm o}$ along the SAW's propagation direction and studied their emission by conventional low temperature $(T= 5\,\mathrm{K})$ microphotoluminescence $(\mu$-PL). For the photogeneration of electron-hole pairs we used a pulsed diode laser $(E_{\mathrm{laser}}=1.88\,{\rm eV})$ which we focused by a 50$\times$ microscope objective to a $\sim2\,\mu$m diameter spot. The emission of the NWs was collected via the same objective, dispersed by a 0.5 m grating monochromator and the signal was detected time integrated by a liquid N$_2$ cooled Si-CCD camera. By setting the frequency of the SAW $f_{\rm SAW}$ to a multiple integer of the repetition frequency of the laser pulses $n\cdot f_{\rm laser}=f_{\rm SAW}$, charge carriers can be generated at a fixed point relative to the SAW. By tuning the delay time $\tau$ between laser and SAW excitation from 0 to $T_{\mathrm{SAW}}$ we are able to pump the NWs at every point of the SAW cycle and, thus resolve the full temporal information of the SAW-driven dynamics \cite{Voelk:11a,Kinzel:11,Fuhrmann:11}.\\

A typical emission spectrum of an individual NW with no SAW applied is plotted in Figure \ref{Fig:1}(c), recorded at low optical pump powers of $P_{\mathrm{laser}}\sim 200\,{\rm nW}$, corresponding to an optical power density of $\sim 6\,{\rm W/cm^2}$. The dominant PL signal centered at $E_{core}=1.525\,{\rm eV}$ can be attributed to carrier recombination in the GaAs core of the NW. We attribute the $\sim 10\,{\rm meV}$ shift with respect to the bulk \ce{GaAs} band gap to strain building up in the NW during cool down due to the largely dissimilar thermal expansion coefficient of \ce{LiNbO3} and \ce{GaAs}. In addition, the core emission exhibits a tail towards lower energies confirming the presence of twin defects\cite{Spirkoska:09}. The PL of the 5nm thick GaAs-QW is shifted to higher energies to $E_{QW}=1.57\,{\rm eV}$ due to quantum confinement. At the highest energies shown here we detect an emission band consisting of a series of single sharp lines. The origin of these interesting features is currently controversially discussed as arising from perfectly ordered and faceted islands \cite{Heiss2013a} or randomly distributed \cite{Rudolph2013} alloy fluctuations and defects within the Al$_{0.3}$Ga$_{0.7}$As shell.
Since the optical excitation occurs at lower energies ($E_{\mathrm{laser}}=1.88\,{\rm eV}$) compared to the band gap of Al$_{0.3}$Ga$_{0.7}$As $(E_{\rm Al_{0.3}Ga_{0.7}As}=1.92-1.96\,{\rm eV})$ carriers are only generated in the GaAs core and QW and in these below-band gap localized QD-like recombination centers. The hierarchy of these energetics, $E_{\rm Al_{0.3}Ga_{0.7}As}>E_{\rm laser}>E_{EC}>E_{QW}>E_{core}$, are included in the schematics in Figure \ref{Fig:1}(b). The \emph{quasi-resonant excitation} conditions will be of great relevance for the interpretation and modeling of our experimental data in the following.
For our experiments presented in this paper, we focus on isolated groups of emission lines at the \emph{low energy} tail of this emission band. An emission band extending to such low energies is observed for the majority of the NWs from this growth with their line intensities varying from NW to NW. Moreover, these energies are compatible with those reported in Ref. \cite{Heiss2013a}. A series of spectra excited at the band edge of the Al$_{0.3}$Ga$_{0.7}As$ barrier of NWs from this growth run and high-resolution transmission electron micrographs (HRTEM) of a reference sample\cite{Funk2013a} are presented in the Supporting Information. These data suggest that enhanced alloy fluctuations in the ${\rm Al_{0.3}Ga_{0.7}As}$ shell may be the origin of the pronounced defect emission band of the NWs studied here\cite{Rudolph2013}. In the spectrum of a single NW shown in Figure \ref{Fig:1}(c) we identify signatures from two individual ECs located within the NW shell. These signals are found at $\sim 1.631\,{\rm eV}$ and $\sim 1.674\,{\rm eV}$ at the low energy side of the Al$_{0.3}$Ga$_{0.7}$As band. In the following we present a detailed study performed on the higher energy EC which we refer to as EC1. A closer examination of the spectrum of EC1 in Figure \ref{Fig:1}(d) reveals that the emission consists of one dominant emission line at 1.6741\,eV which we attribute to recombination of the charge neutral single exciton ($1X^0={\rm 1e+1h}$), consisting of a single electron (e) and a single hole (h). The two weaker emission lines at 1.6717\,eV and 1.6701\,eV arise from a charged exciton ($1X^*$) with a dissimilar number of electrons and holes and the neutral biexciton ($2X^0={\rm 2e+2h}$), respectively. From these spectral shifts we obtain a biexciton binding energy of $\sim 4\,{\rm meV}$ and a renormalization energy of the observed charged exciton of $\sim 2.4\,{\rm meV}$.
This line assignment is further confirmed by laser excitation power dependent spectroscopy. Emission spectra of EC1 recorded for $P_{\mathrm{Laser}}$ ranging between 10 and 800\,nW are plotted in Figure \ref{Fig:1}(d). While three emission lines show a clear increase of intensity at low optical pump powers, $1X^0$ and $1X^*$ saturate at the highest power levels, in strong contrast to $2X^0$. This behavior becomes even clearer in the extracted peak intensities, which are plotted as a function of $P_{\mathrm{Laser}}$ in double-logarithmic representation in Figure \ref{Fig:1}(e). From the observed slopes in this representation we identify different power-law dependencies $(I\propto P_{\rm Laser}^m)$ for the three emission lines. For $1X^0$ and $2X^0$ we find exponents of $m=0.9\pm 0.1$ and $m=1.8\pm0.1$, respectively, close to the expected linear $(m=1)$ and quadratic $(m=2)$ dependencies \cite{Brunner:94a}. 
We want to note at this point, that this assignment is based on a model originally established for planar heterostructure QDs. One striking property of the ECs studied here is their low measured ground state transitions energy. Simply assuming the EC as a cube of \ce{GaAs} in \ce{Al_{0.3}Ga_{0.7}As} heterostructure QD, the measured confinement energy of $~150\,{\rm meV}$ would imply a QD size of $\sim 8.3\,{\rm nm}\times8.3\,{\rm nm}\times8.3\,{\rm nm}$. Such large \ce{GaAs} inclusions have not been observed in structural characterization on reference NWs grown under identical conditions \cite{Rudolph2013}. Therefore we conclude that the system studied here is of more complex nature. Nevertheless, the cubic heterostructure QD defines upper boundaries for energy barriers which we use to model our experimental data.\\

\begin{figure}[htb]
	\begin{center}
		\includegraphics[width=0.85\columnwidth]{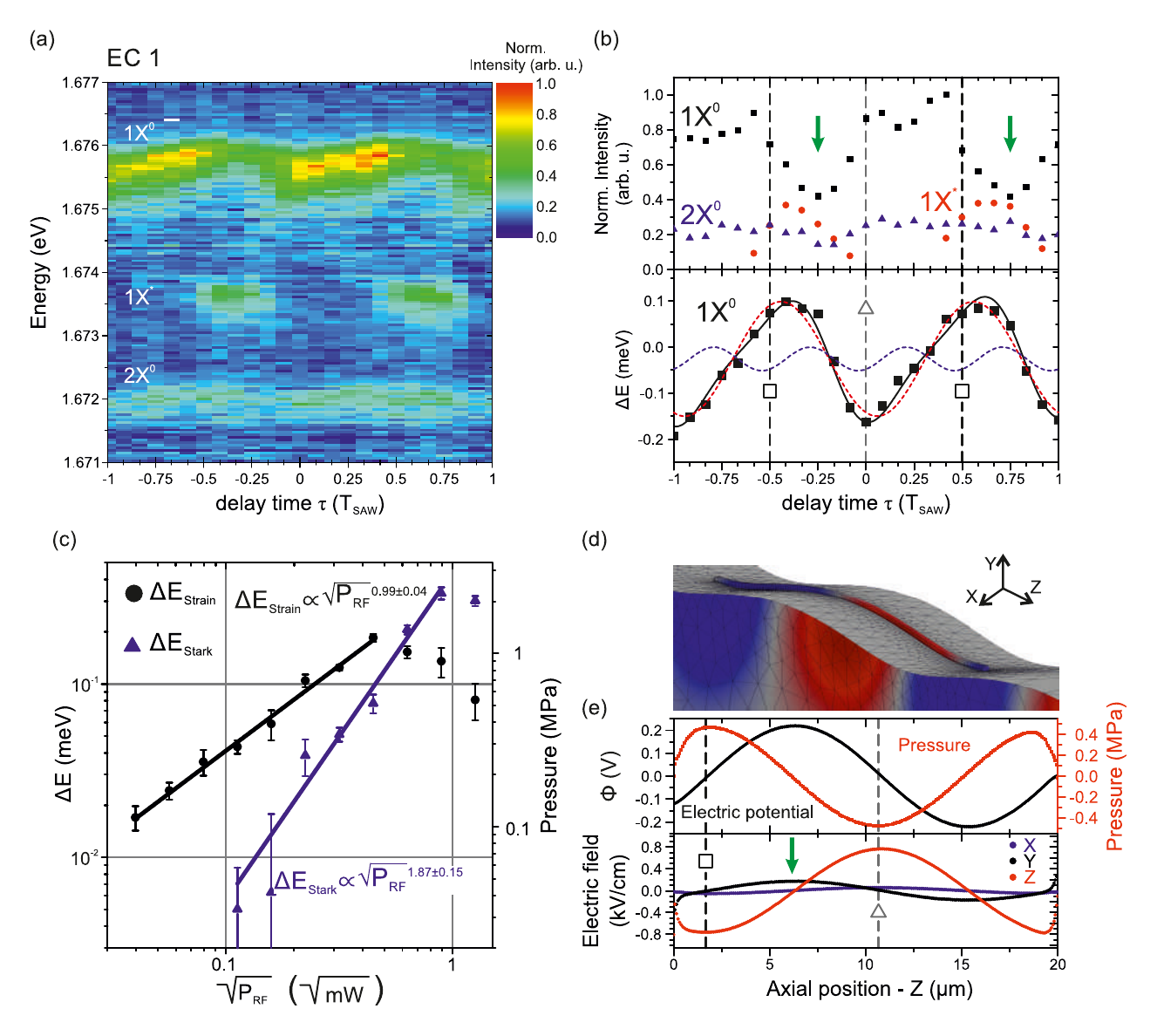}
		\caption{\textbf{SAW tuning of EC emission and FE simulations} -- (a) Stroboscopic PL spectra of EC1 recorded over two acoustic cycles $(P_{\rm RF}=-10\,{\rm dBm})$ showing spectral and anti-correlated intensity oscillations due to dynamic SAW tuning. (b) Extracted normalized intensities of the different QD lines (upper panel) and spectral modulations (lower panel, symbols) extracted from the data shown in (a). The full line in the lower panel is a best fit of Equation 2 to the data. The broken red and blue lines indicated the fitted contributions of ${\Delta}E_{\rm Strain}$ and ${\Delta}E_{\rm Stark}$, respectively. (c) $\Delta E_{strain},\:(\bullet)$ and $\Delta E_{Stark},\:(\blacktriangle)$ as a function of $\sqrt{P_{\rm RF}}\propto A_{\rm SAW}$ in double-logarithmic representation. Lines are power-law fits to the experimental data reproducing the experimental $\Delta E_{strain}\propto A_{\rm SAW}$ and $\Delta E_{Stark} \propto A_{\rm SAW}^2$ dependencies. (d) Displacement and electric potential (color code) of a \ce{GaAs} NW on YZ-\ce{LiNbO3} hybrid calculated by FE modeling. (e) Extracted electric potential (upper panel, black) and hydrostatic pressure (upper panel, red) and electric field components (lower panel) in the center of the NW. The maximum and minimum of $F_{\rm Z}$ are indicated by vertical lines and corresponding $(\triangle)$ and $(\Box)$ in (b) and (e).}
	\label{Fig:2}
	\end{center}
\end{figure}

Following this characterization of the unperturbed EC emission we now turn to its control by a SAW as shown in the schematic of Figure \ref{Fig:1}(a). We study the emission of EC1 with a SAW generated by applying a resonant RF signal to the IDT. In Figure \ref{Fig:2}(a) we present stroboscopic emission spectra of EC1 for $P_\mathrm{\rm RF}=-10\,\mathrm{dBm}$ which are plotted in false-color representation. As we tune the delay time $\tau$ over two full acoustic cycles, we resolve \emph{both} pronounced intensity and spectral modulations of the three PL lines. Both effects exhibit a clear dependence on the time delay $\tau$. The intensity oscillations between $1X^0$ and $1X^*$ show clear anti-correlation which becomes more clearly visible in the extracted peak intensities of the three emission lines in the upper panel of Figure \ref{Fig:2} (b). The observed anti-correlation between different charge configurations indicates that the moment of excitation during the acoustic cycle, $\tau$, programs the charge state of the EC, similar to our previous experiments in planar heterostructure systems\cite{Voelk:11a,Schulein2013}. 
Before we address the mechanism giving rise to these anti-correlated intensity oscillations we start by an analysis of the spectral tuning. In order to quantify this effect, we extract the energetic shift $\Delta E$ of $1X^0$ which is plotted as symbols in the lower panel of Figure \ref{Fig:2}(d) and exhibits a total modulation bandwidth of $\pm 0.2\,{\rm meV}$. Most interestingly, the modulation itself is a clear superposition of two oscillations, the first following the SAW periodicity and a second exhibiting two oscillations per SAW cycle. This indicates the presence of two couplings, that are dynamically driven by the SAW. On the one hand the SAW induces a dynamic strain field which gives rise to a spectral shift $\Delta E\mathrm{_{Strain}}$ via deformation potential coupling. This contribution has been previously observed for embedded heterostructure QWs \cite{Sogawa:01a} and QDs\cite{Gell:08,*Metcalfe:10,Schulein2013}. Its amplitude $\Delta E_{\rm Strain}$ scales linearly with the hydrostatic pressure $\propto  p$ induced by the SAW. The latter also scales linearly with $\propto A_{\rm SAW}$ and in turn leads to one oscillation per acoustic cycle for this contribution. On the other hand, the SAW-induced electric field $F\propto A_{\rm SAW}$ in the \ce{GaAs} NW leads to a second contribution to the spectral shift via the quantum confined Stark effect (QCSE)\cite{Santos2004}

\begin{equation}
	\Delta E_{\rm Stark}=-p_X F =-\beta F^2.
	\label{QCSE}
\end{equation}

In this equation $\beta$ denotes the polarizability of the exciton and $p_X=\beta F$ is the exciton's electrostatic dipole moment at given $F$. Moreover, the exciton is considered as a classical electrostatic dipole, $p_X=er_{\rm eh}$, with $e$ being the elementary charge and $r_{\rm eh}$ the spatial separation between the centers of gravity of the e and h wave functions. Since $\Delta E_{\rm Stark}=-\beta F^2$, this contribution \emph{always} reduces the emission energy. Moreover, this reduction is maximum at the two distinct $\tau$ of maximum and minimum $F$. Therefore, the contribution of the QCSE is expected to lead to an oscillation with angular frequency $2\omega_{\rm SAW}$.
Taken together, since both strain and electric fields scale linear with the acoustic amplitude $A_{\rm SAW}$, we expect that $\Delta E\mathrm{_{Strain}}\propto A_{\rm SAW}$ and $\Delta E\mathrm{_{Stark}}\propto A_{\rm SAW}^2$.
To discriminate between these two contributions we fit our experimental data by a superposition of two sinusoidal oscillations of angular frequency $\omega_{\rm SAW}$ for the strain tuning and $2\omega_{\rm SAW}$ for the QCSE:

\begin{equation}
	\label{fit}
	\Delta E(\tau)={\Delta}E_{\rm Strain}\cdot\sin\left(\omega_{\rm SAW}\tau\right)+{\Delta}E_{\rm Stark}/2\cdot \sin\left(2 \omega_{\rm SAW}\tau\right).
\end{equation}

From fitting Equation 2 we obtain the total emission energy and the individual contributions ${\Delta}E_{\rm Strain}$ and ${\Delta}E_{\rm Stark}$, which are plotted in lower panel of Figure \ref{Fig:2}(b) as the full black and the broken red and blue lines, respectively. In order to confirm the anticipated power law dependencies $A_{\rm SAW}^n$, we fit Equation 2 to the spectral tuning of EC1 for different $P_{\rm RF}$. The extracted amplitudes $\Delta E_{\rm Strain}$ (black symbols) and $\Delta E_{\rm Stark}$ (blue symbols) are plotted in a double-logarithmic representation as a function of $\sqrt{P_{\rm RF}}$ in Figure \ref{Fig:2}(c). Since $A_{\rm SAW}\propto \sqrt{P_{\rm RF}}$ we expect $n=1$ for $\Delta E_{\rm Strain}$ and $n=2$ for $\Delta E_{\rm Stark}$. Both values are clearly confirmed within the experimental error by linear fits plotted as solid lines in Figure \ref{Fig:2}(c) yielding $n=0.99\pm0.04$ for $\Delta E_{\rm Strain}$ and $n=1.9\pm0.15$ for $\Delta E_{\rm Stark}$, respectively. $\Delta E_{\rm Strain}$ decreases at high acoustic amplitudes which points to a partial detachment at large $P_{\rm RF}$. Moreover, we convert $\Delta E_{\rm Strain}$ to a hydrostatic pressure using the deformation potential induced bandgap variation for $\rm Al_{0.22}Ga_{0.78}As$ of $\frac{dE_{g}}{dp}=150\frac{\mu{\rm eV}}{\rm MPa}$\cite{
Qiang1990}. The such obtained hydrostatic pressure is given on the right axis of Figure \ref{Fig:2}(c).\\

To quantify these experimental observations, we performed a finite element (FE) modeling of the interaction between the acoustic and piezoelectric fields of the SAW on the \ce{LiNbO3} substrate and the \ce{GaAs} NW. In these simulations we assumed a $280{\,\rm nm}$ diameter $(111)$-oriented NW with $\{110\}$ facets and increased the NW length to $20\,\mu{\rm  m}>\lambda_{\rm SAW}=18\,\mu{\rm m}$ to calculate all relevant parameters in a single simulation. As in our experiments, the axis of the NW is aligned with the Z-propagating SAW which is excited by $P_\mathrm{\rm RF}=-10\,\mathrm{dBm}$. Figure \ref{Fig:2} (d) shows the calculated structural deformation (enhanced by a factor of $\sim 5\cdot10^4$) and electric potential, $\Phi,$ (color coded). Our FE simulation clearly demonstrates that both mechanical and electric excitation in the \ce{LiNbO3} substrate are coupled into the \ce{GaAs} NW. Furthermore, we extracted the hydrostatic pressure $p$, the electric potential $\Phi$ plotted in red and black in the the upper panel of Figure \ref{Fig:2}(e) as well as the longitudinal ($F_{\rm Z}$, red) and transverse ($F_{\rm X}$, blue; $F_{\rm Y}$, black) components of the electric field in the NW in the lower panel. Of these components, the longitudinal $F_{\rm Z}$ component, is dominant and in addition to the expected $F_{\rm Y}$ a second, smaller transverse component $F_{\rm X}$ is induced due to a structural deformation of the NW induced by piezomechanical coupling. For our YZ-\ce{LiNbO3}, the oscillation of $p$ is phase-shifted by $\pi/2$ and $\pi$ with respect to the oscillation of the transverse $F_{\rm Y}$ and longitudinal $F_{\rm Z}$ components, respectively. At one distinct phase during the SAW oscillation, the pressure, $p$, is maximum negative (tensile) and the longitudinal field component, $F_{\rm Z}$, is maximum positive. At this particular local phase both contributions reduce the EC emission energy and give rise to its absolute minimum. We identify this absolute minimum in the stroboscopic PL data and assign it to $\tau = 0$. This calibration is indicated by a vertical dashed line and $(\triangle)$ in Figure \ref{Fig:2} (b) and (e). In turn, this implies that $p$ and $F_{\rm Z}$ are maximum positive (compressive) and negative, respectively at $\tau =\pm T_{\rm SAW}/2$ as marked by vertical dashed lines and $(\Box)$ in Figure \ref{Fig:2} (b) and (e). The calculated hydrostatic pressure of $p_{FE}=0.45\,{\rm MPa}$ is $\sim 35\%$ smaller than $p_{exp}=0.7\,{\rm MPa}$ extracted from the experimental data using a simple hydrostatic model. This discrepancy might arise from limitations in the conversion of experimental parameter $P_{\rm RF}$ to the simulation parameters, the large variations of reported deformation potential couplings in particular of \ce{(Al)GaAs}\cite{Qiang1990,Pollak1968,*Vurgaftman:01} and the hydrostatic approximation neglecting contribution of off-diagonal strain components. From the amplitude of the QCSE oscillation given by Equation 1, we can determine the e-h distance $r_{\rm eh}=\Delta E_{\rm Stark}/e|F|$. Taking into account that $|F|\sim F_{\rm Z}$ for our NW we can estimate $r_{eh}= 1.5\pm 0.2\,{\rm nm}$ as the e-h separation for $P_\mathrm{RF}=-1\,\mathrm{dBm}$ at which we observe the maximum of $\Delta E_{\rm Stark}$. Since the magnitude of the Stark shift reflects the width of a nanostructure\cite{Polland:85} we further conclude that $r_{eh}= 1.5\pm 0.2\,{\rm nm}$ provides a measure for both the e-h separation and the size of the emission center.\\

\begin{figure}[htb]
	\begin{center}
		\includegraphics[width=0.3\columnwidth]{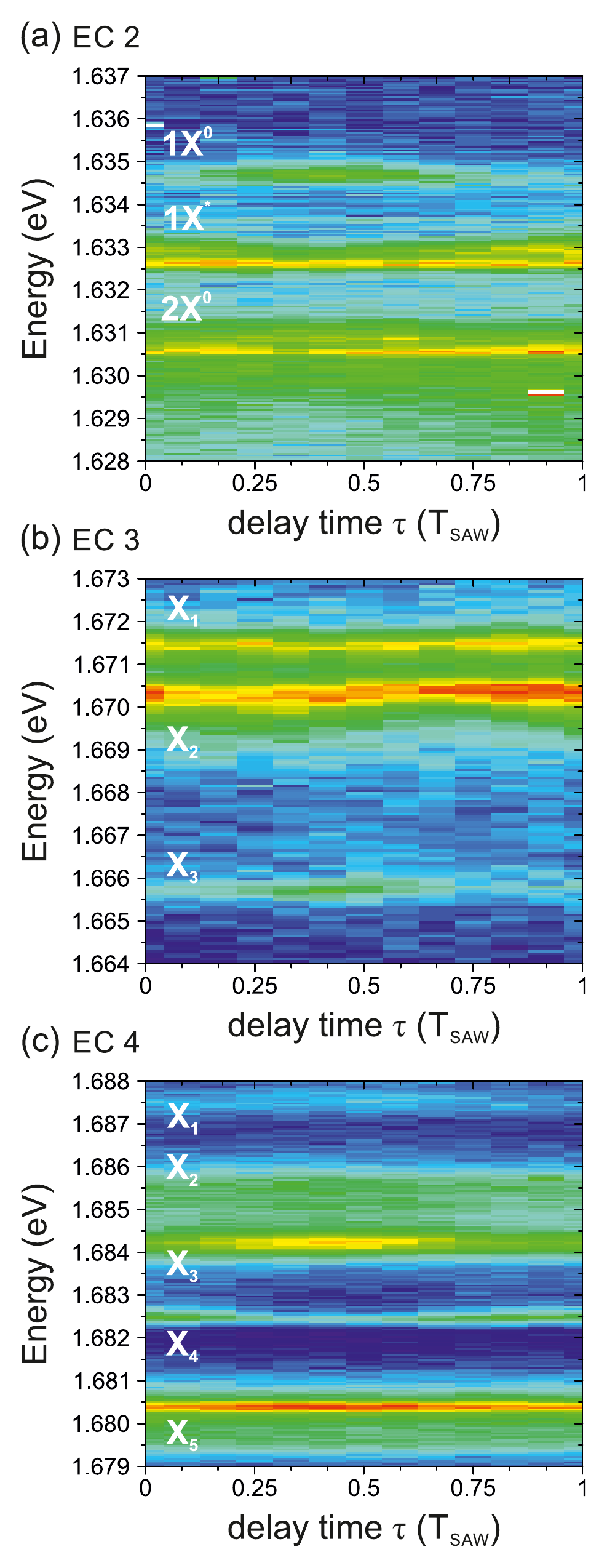}
		\caption{\textbf{Anti-correlated intensity oscillations as a general fingerprint} -- Stroboscopic PL spectra plotted over one acoustic cycle for (a) EC2 located on the same NW as EC1, (b) EC3 located in a different NW on the same SAW-chip and (c) EC4 located in a different NW on a different SAW-chip. The different modulation contrast indicates different efficiencies of the underlying tunneling mechanism for the three ECs. The color scale is the same as in Figure \ref{Fig:2} (a).}
		\label{Fig:3}
	\end{center}
\end{figure}

Finally we address the anti-correlated intensity modulation observed in the experimental data and develop a model to describe their microscopic origin. In Figure \ref{Fig:3}(a-c) we present stroboscopic PL spectra of three different ECs, labeled EC2, EC3 and EC4. EC1-EC3 are located in two different NWs on the same substrate. EC4 is located in an third NW on a different SAW chip. All presented data were recorded at identical RF power $P_\mathrm{RF}=-10\,\mathrm{dBm}$. EC2 shows an excitation power dependence similar to EC1, therefore we analogously assign the observed emission lines to $1X^0$, $1X^*$ and $2X^0$ from high to low energies. In contrast, the same type of data from EC3 and EC4 are not conclusive and, consequently, we instead label the observed emission lines $X_1$-$X_3$ and $X_1$-$X_5$, respectively. A comparison of the $\tau$-dependent evolution of the emission signals of the four ECs clearly shows that the anti-correlated intensity oscillations between different exciton transitions seem indeed to be a general fingerprint for SAW response of the optical emission of these types of ECs. Most strikingly, the modulation contrast differs strongly from dot to dot as it is less developed for EC2, EC3 and EC4 compared to EC1. This points towards the fact that the underlying mechanism is sensitive to the QDs/ECs properties and/or environment, in strong contrast to planar, embedded QD nanostructures \cite{Voelk:11a,Voelk:12,Schulein2013} for which the SAW control of the QD occupancy state is highly reproducible from dot to dot. This reproducibility furthermore confirms similar coupling of the SAW to the NW.\\

\begin{figure}[htb]
	\begin{center}
		\includegraphics[width=0.75\columnwidth]{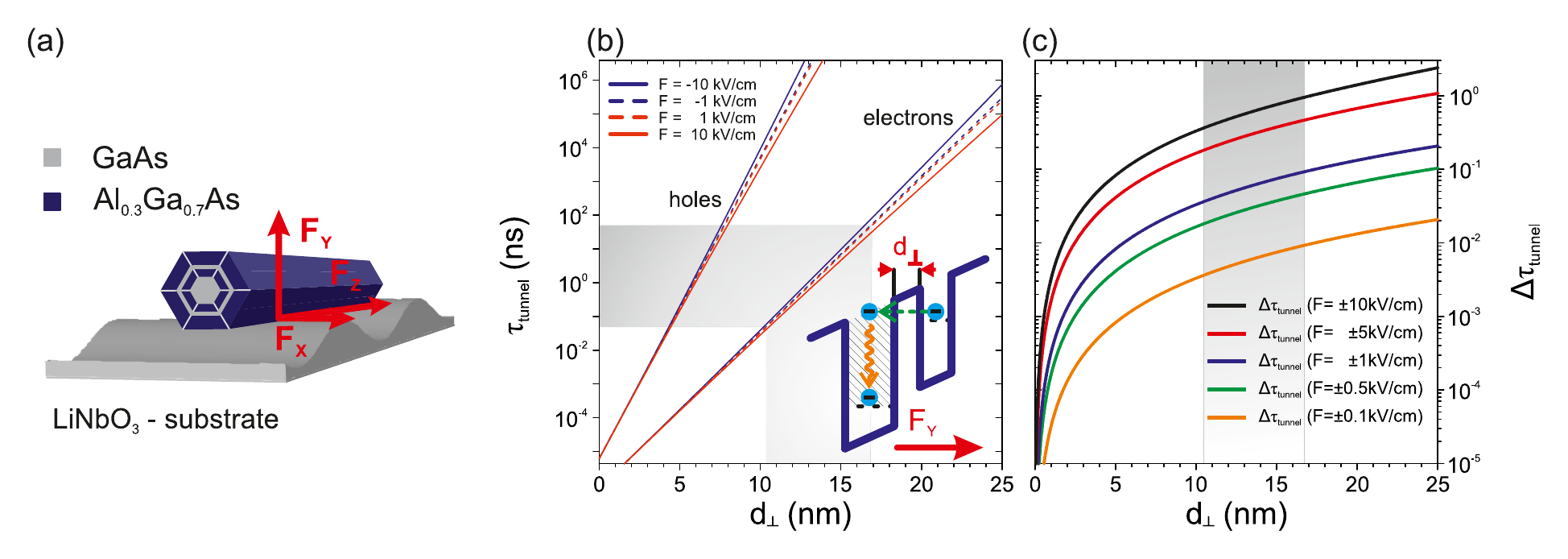}
		\caption{\textbf{WKB modeling of SAW-controlled tunneling} -- (a) Schematic of relative orientation of NW and the components of the SAW-induced electric fields. (b) Tunneling time for e's and h's for different electric fields as a function of barrier thickness calculated using Equation 3. The shaded area indicates the range of distances for which $50\,{\rm ps}\leq \tau_{tunnel,e}\leq50\,{\rm ns}$ is in the range of PL decay times. A schematic bandstructure and the underlying process is shown as an inset. (c) Calculated relative modulation of the tunneling time given by Equation 4 predicting modulations of a few percent for experimentally accessible SAW-induced electric fields.
		\label{Fig:4}}
	\end{center}
\end{figure}

Taken together, the microscopic origin of the observed intensity oscillations reported here is fundamentally different to that observed for embedded QD nanostructures. This established mechanism relies on the photogeneration in a continuum of states where the longitudinal electric field of the SAW induces spatio-temporal carrier dynamics. These dynamics in turn lead to an acoustically regulated injection of e's and h's into the energetically lower QD states. We can exclude this mechanism as the origin of the intensity oscillations observed for our NW ECs for three reasons. First, considering the energetic ordering of the effective band gaps of the $E_{\rm Al_{0.3}Ga_{0.7}As}>E_{\rm laser}>E_{QD}E_{EC}>E_{QW}>E_{core}$ (see Figure \ref{Fig:1}(b)), SAW-driven injection can only occur from the AlGaAs shell. In addition, no free carriers are photogenerated in the shell which could be injected into the ECs. Moreover, at the low acoustic powers applied \emph{no pronounced signatures for SAW-driven spatio-temporal carrier dynamics} are observed for both the \ce{GaAs} core and the QW as demonstrated in the Supporting Information. Since no free carriers can be injected into the EC by the SAW, the mechanism underlying the observed intensity oscillations has to rely on a \emph{SAW-mediated carrier extraction}. Since all experiments are performed at low temperatures, we attribute our observation as arising from tunnelling of e's from the EC to the 3D and 2D continuum states of the \ce{GaAs} core, capping and QW which is modulated by the SAW-induced electric fields. Due to the alignment of the NW with respect to the SAW propagation the longitudinal component $F_{\rm Z}$ is oriented along the NW axis and thus the radial heterostructure. In contrast, the transverse components $F_{\rm X}$ and $F_{\rm Y}$ are oriented perpendicular to the interfaces of the radial heterostructure. The relative alignments are depicted schematically in Figure \ref{Fig:4}(a). Since ECs are embedded in the \ce{AlGaAs} shell, $F_{\rm X}$ and $F_{\rm Y}$ but not $F_{\rm Z}$ can modify the tunneling of carriers from their confined energy levels to the \ce{GaAs} core, capping and QW. Accounting for $F_{\rm Y} \gg F_{\rm X}$ we expect only a minor contribution of $F_{\rm X}$ which we neglect in the following. $F_{\rm Y}$ oscillates with an amplitude $F_{{\rm Y},max}$ over one acoustic cycle and thus periodically lowers and raises the tunneling barrier between the EC and the continuum. Since this process is directional, it manifests itself by an increased tunneling probability for $F$ antiparallel to the tunnel direction as shown in the inset of Figure \ref{Fig:4} (b). This in turn gives rise to a single intensity oscillation per acousic cycle as observed for all four ECs.
A direct comparison of the intensity and spectral oscillations of EC1 in Figure \ref{Fig:2} (b) clearly shows that the reduction of the $1X^0$ and increase of the $1X^*$ signals occur for $-T_{\rm SAW}/2\leq \tau \leq 0$. In this time interval $F_{\rm Y}$ is positive and points upwards in $+ \rm Y$-direction which directly reflects the tunneling direction of the electron. This correlation is indicated by the green arrows in Figure \ref{Fig:2} (b) and (e). In the time interval $0\leq \tau \leq +T_{\rm SAW}/2$ no $1X^*$ emission is detected since the antiparallel alignment $F_{\rm Y}$ and the tunneling direction suppresses the carrier extraction. A comparison of the spectral and intensity oscillations of the dominant  emission lines of EC1 $(1X^0)$ and EC3 $(X_2)$ in Figures \ref{Fig:2} (a,b) and \ref{Fig:3} (b) provide a further point of evidence. While for $1X^0$ of  the minimum intensity occurs at the steeper, falling edge of the spectral modulations, the situation is reversed for $X_2$ of EC3, which exhibits its maximum intensity at this time during the acoustic cycle.\\

We quantify the efficiency of this process and its control by $F_{\rm Y}$ by performing Wentzel-Kramers-Brillouin (WKB) calculations of the tunneling time, $\tau_{tunnel}$ as function of the EC-continuum separation $d_{\perp}$. This approach has been established to quantify carrier tunneling from planar QD system through a triangular barrier (Fowler-Nordheim tunneling) at high $F_{\rm Y}$ \cite{Fry:00b,*Krenner:08a,*Mueller:12a}. However, for our ECs tunneling occurs through a rectangular barrier as indicated in the inset of Figure \ref{Fig:4}(b). For this barrier we obtain for the tunneling rates for e's and h's as a function of $F_{\rm Y}$ \cite{Schuegraf1992} 

\begin{equation}
\tau_{tunnel,e/h}^{-1}=\frac{\hbar\pi}{2m_{e/h}^*L^2}\cdot\exp\left[\frac{-4\sqrt{2m_{e/h}^*E_{i,e/h}^3}}{3\hbar eF_{\rm Y}}\cdot\left(1-\left(1-\frac{F_{\rm Y}\cdot d_\perp}{E_{i,e/h}}\right)^{\frac{3}{2}}\right)\right].
\label{tunnel}
\end{equation}

We evaluate Equation 3 as a function of the barrier thickness $d_{\perp}$, dot size $L=r_{eh}= 1.5\,{\rm nm}$ and  barrier heights and effective masses of $E_{i,e}=160\,\mathrm{meV}$, $m_e^*=0.067\mathrm{m}_0$ and $E_{i,h}=80\mathrm{\,meV}$, $m_{h}^*=0.5\mathrm{m}_0$ for e's and h's, respectively. These values correspond to a pure \ce{GaAs} heterostructure QD in a \ce{Al_{0.3}Ga_{0.7}As} matrix. This type of QD is prototypical which we confirmed by a detailed investigation of the impact of different parameters in our WKB model presented in the Supporting Information of this letter. In Figure \ref{Fig:4}(b) we plot the results for both carrier species for moderate, $F_{\rm Y}=\pm1\,{\rm kV/cm}$ (dashed lines) and high $F_{\rm Y}=\pm10\,{\rm kV/cm}$ (solid lines) as a function of $d_{\perp}$. Our WKB calculations confirm that the different effective masses favors the tunneling of e's $(\tau_{tunnel,e}\ll\tau_{tunnel,h})$ and we consequently identify it as the underlying carrier extraction mechanism. These calculations predict that $\tau_{tunnel,e}\leq50\,{\rm ps}$ for separation $d_{\perp,crit}\leq 10.5\,{\rm nm}$. Such fast tunneling processes efficiently depopulate the EC on timescales faster than typical radiative lifetimes\cite{Heiss2013a} which strongly suppresses its PL efficiency. A similar reasoning can be applied in the limit of large separations. For $d_{\perp}\geq 17\,{\rm nm}$, $\tau_{tunnel,e}\geq50\,{\rm ns}$ which does not allow for efficient carrier extraction within excitonic radiative lifetimes. This range of times and the corresponding distances are marked by the shaded areas in Figure \ref{Fig:4}. For larger separations, tunneling still occurs with low probability during the radiative lifetime. The long tunneling times in turn give rise to a  build up of charge which manifests itself in a reduction of the modulation contrast and a multiplet of emission lines as observed for EC4.  From this we conclude, that for tunneling occuring on the timescales comparable or slower than radiative processes, any type of QD or EC of similar confinement in the \ce{AlGaAs} shell has to be separated by $d_{\perp,crit}\geq 10.5\,{\rm nm}$ from the QW, NW core or the GaAs capping. 
As shown in the Supplementary Information this critical distance reduces slightly to $d_{\perp,crit}\sim 7\,{\rm nm}$ for the maximum condution band offset occuring for a \ce{Al_{0.45}Ga_{0.55}As}-\ce{GaAs} interface. These lengthscales are fully compatible with the nominal \ce{Al_{0.3}Ga_{0.7}Ga} barrier thicknesses in the radial heterostructure of our NWs.\\

The SAW modulates $F_{\rm Y}$ between $\pm F_{{\rm Y},max}$ over one acoustic period and gives rise to a dynamic modulation of the tunneling time. We quantify the amplitude of this modulation by calculating the dimension-less relative variation of  $\tau_{tunnel}$ for switching between $\pm F_{\mathrm{Y},max}$ relative to $\tau_{tunnel}(F=0)$,

\begin{equation}
	\label{rel}
	\Delta\tau_{tunnel}=\frac{|\tau_{tunnel}(-F_{{\rm Y},max})-\tau_{tunnel}(+F_{{\rm Y},max})|}{\tau_{tunnel}(F_{\rm Y}=0)}.
\end{equation}

We plot the $d_{\perp}$-dependence of $\Delta \tau_{tunnel}$ for different $F_{{\rm Y},max}$ in Figure \ref{Fig:4}(c). The solution of Equation 4 show that for a constant $F_{{\rm Y},max}$ a monotonic increase of $\Delta \tau_{tunnel}$ with increasing barrier thickness, which saturates for $d_{\perp}> 10\,{\rm nm}$. Most importantly, in the electric field range accessible by a SAW, $|F_{{\rm Y},max}|\leq 10 \, {\rm kV/cm}$, we obtain values $10^{-3}\leq\Delta \tau_{tunnel}\leq 0.7$ in the range of distances for which $\tau_{tunnel}$ can modulate radiative processes. In the experimental data presented in Figures \ref{Fig:2} and \ref{Fig:3}, intensity oscillations are driven by the larger transverse component $F_{{\rm Y},max}\sim 0.2-0.3\,{\rm kV/cm}$. For such field amplitudes, Equation 4 predicts $\Delta \tau_{tunnel}$ between $1\%$ and $5\%$. The observed anti-correlated intensity oscillations exhibit a similar contrast and, thus confirm our identification of SAW-controlled tunneling as the underlying mechanism.\\

Finally we want to discuss implications of our observations on the nature of the QD-like emission. The first striking property of the ECs studied here is their low measured ground state transitions energy. Since the emission of some of the ECs studied here exhibit the expected excitation power dependence, e.g. EC1, a confining potential for at least one carrier species has to be present which gives rise to the different occupancy states. The QD-like properties could arise from a combination of quantum confinement of radial alloy fluctuations and point defects\cite{Rudolph2013}. Moreover, occasional twin defects occurring in the NW core can extend into the radial heterostructure\cite{Algra2011} and could lead to an additional but weak modulation of the band edges. The results of our WKB modeling suggest, that the ECs studied are at minimum distance of $d_{\perp,crit}\geq 10.5\,{\rm nm}$ from a continuum. For all QDs studied so far, we observe SAW-driven intensity modulations, however the contrast of these oscillations differs largely from EC to EC. The latter finding in turn implies different efficiencies of the underlying tunneling mechanism. Such different efficiencies suggest a broad and random distribution of $d_{\perp}$ in our sample rather than a high level of spatial ordering. An expanded discussion of our WKB modeling for alternative QD morphologies can be found in the Supporting Information of this letter.\\

To summarize, we investigated the optical properties of QD-like emission centers forming in \ce{Al_{0.3}Ga_{0.7}As} layers of radial heterostructure NWs and their dynamic control by a SAW.
The implications of our findings are threefold. First, we demonstrated that the emission of these centers in our sample can exhibit QD-like properties, in particular few-particle shell filling which we attribute to a combination of radial alloy fluctuations and point defects in the \ce{Al_{0.3}Ga_{0.7}As} layers. Second, in our SAW experiments we demonstrated for the first time spectral oscillations of the EC emission by \emph{both} SAW induced strain and electric fields. These spectral oscillations are accompanied by pronounced intensity oscillations driven by SAW-controlled carrier extraction from the EC to a continuum of higher dimensionality in the heterostructure. By comparing our data to numerical simulations we identify quantum tunneling as the underlying mechanism. Our WKB-simulations suggest, that the emission centers in our system are randomly distributed in the \ce{Al_{0.3}Ga_{0.7}As} shell at a minimum separation of $d_{\perp,crit}\geq 10.5\,{\rm nm}$. This mechanism has a third important consequence. In all previously studied QD systems, such intensity oscillations have been driven by acoustically regulated carrier injection \cite{Schulein2013}. Here, we experimentally demonstrated SAW-controlled extraction of carriers from an optically active QD within its radiative lifetime into a system of higher dimensionality. This opens the possibility to combine approaches based of acoustic charge conveyance \cite{Rotter:99a,*Hermelin:11,*McNeil:11} on contacted single NWs. Such systems are currently already within reach using an axial heterostructure NW architecture\cite{Kouwen:10a,*Reimer2011}.

\subsection*{Author information}
The authors declare no competing financial interest.

\begin{acknowledgement}
This work was financially supported by the Deutsche Forschungsgemeinschaft (DFG) via Sonderforschungsbereich SFB 631 (Projects B1 and B5) and the Emmy Noether Program (KR3790/2-1) and by the European Union via SOLID and the FP7 Marie-Curie Reintegration Grant.
\end{acknowledgement}

\begin{suppinfo}
(i) Emission spectra from typical NWs from this growth run excited above the \ce{Al_{0.3}Ga_{0.7}As} bandgap. (ii) Cross-sectional HRTEM of a reference sample\cite{Funk2013a}. (iii) PL suppression by SAW of the GaAs core and QW emissions. (iv) Details on WKB modeling for different input parameters corresponding to alternative QD morphologies\cite{Heiss2013a}.
\end{suppinfo}


\providecommand*\mcitethebibliography{\thebibliography}
\csname @ifundefined\endcsname{endmcitethebibliography}
  {\let\endmcitethebibliography\endthebibliography}{}

\end{document}